\documentclass[a4paper,twocolumn]{article}
\usepackage{rsj2024e}
\usepackage{url}
\usepackage{tikz}

\usepackage{cite}
\usepackage{amsmath}
\usepackage{amsfonts}
\usepackage{array}
\usepackage{caption}
\usepackage{colortbl}
\usepackage{booktabs}
\usepackage{algorithm}
\usepackage{float}
\usepackage{algorithmic}
\usepackage{graphicx}

\begin{document}
\title{Interactive Holographic Visualization for 3D Facial Avatar
}
\author{*Tri Tung Nguyen NGUYEN (Ritsumeikan University), Yasuyuki FUJII (Ritsumeikan University), Dinh Tuan TRAN (Ritsumeikan University), Joo-Ho LEE (Ritsumeikan University)}

\abstract{Traditional methods for visualizing dynamic human expressions, particularly in medical training, often rely on flat-screen displays or static mannequins, which have proven inefficient for realistic simulation. In response, we propose a platform that leverages a 3D interactive facial avatar capable of displaying non-verbal feedback, including pain signals. This avatar is projected onto a stereoscopic, view-dependent 3D display, offering a more immersive and realistic simulated patient experience for pain assessment practice. However, there is no existing solution that dynamically predicts and projects interactive 3D facial avatars in real-time. To overcome this, we emphasize the need for a 3D display projection system that can project the facial avatar holographically, allowing users to interact with the avatar from any viewpoint. By incorporating 3D Gaussian Splatting (3DGS) and real-time view-dependent calibration, we significantly improve the training environment for accurate pain recognition and assessment.
% \\[1em]
% \includegraphics[width=1\textwidth, trim={0cm 2.5cm 0cm 2.5cm}, clip]{diagrams/overview.pdf}
% \captionof{figure}{\textbf{3D Holographic Avatar Projection Overview}. A novel pipeline combining 3D Gaussian Splatting and light-field shading enables real-time, photo-realistic facial avatar hologram animations, streamed to the 3D Looking Glass display. The OpenGL-based light-field shading program uses FLAME expression parameters with its 3D Gaussian primitive correspondence and the display calibration parameters to render a multi-view quilt that adapts to the audience's viewpoint.}
}

\setlength{\baselineskip}{4.4mm}
\maketitle

\thispagestyle{empty}
\pagestyle{empty}

% \twocolumn[
% \begin{figure*}[H]
% \centering
% \label{fig:data_flow}

% \caption{The system delivers a photo-realistic 3D facial avatar to the user via a stereoscopic multi-view display. The display projects different discrete views in real-time based on the audience's perspective.}
% \end{figure*}
% ]

\section{Introduction}

\begin{figure*}[t!]
\includegraphics[width=1\textwidth, trim={0cm 2.5cm 0cm 2.5cm}, clip]{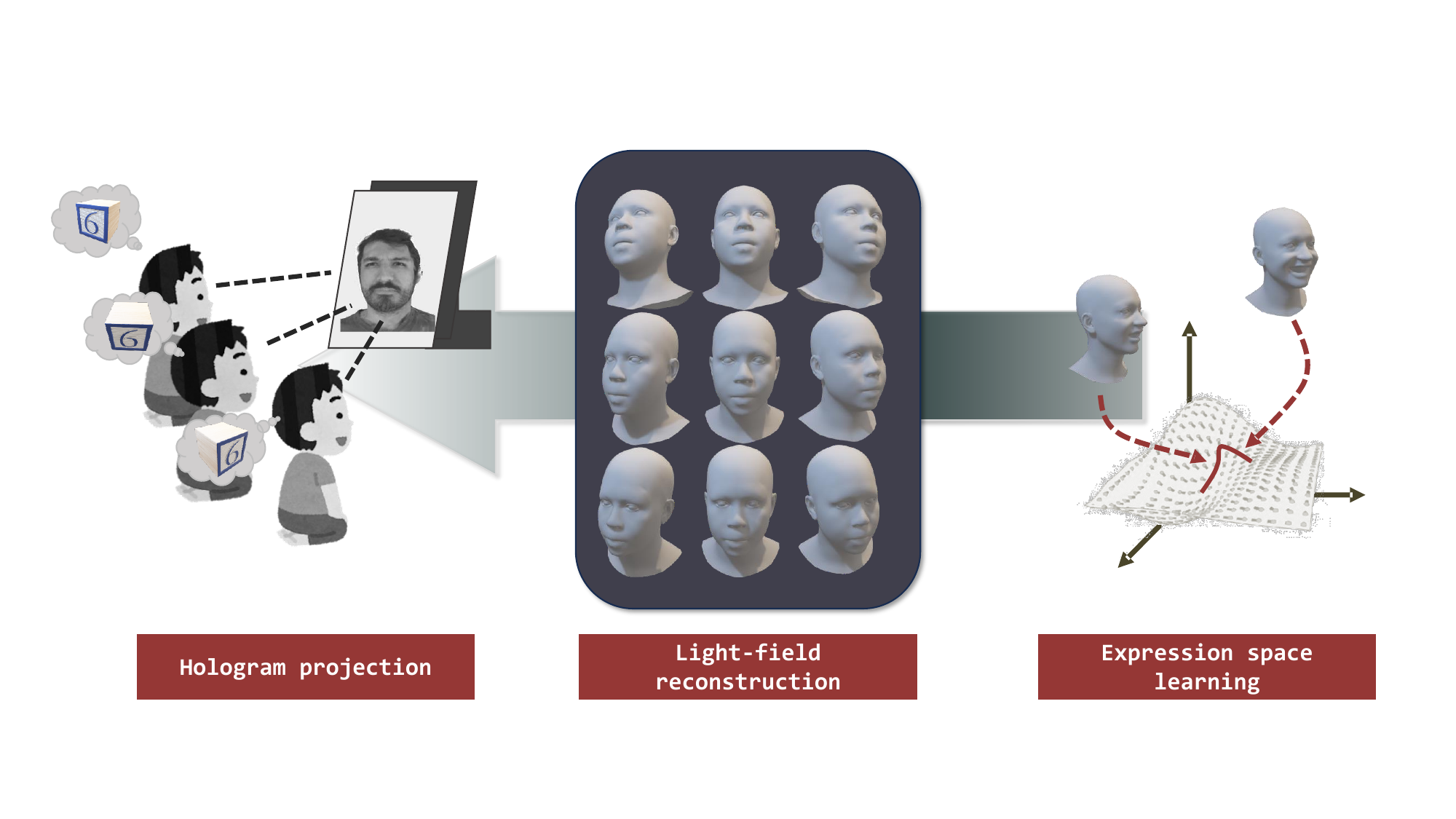}
\captionof{figure}{\textbf{3D Holographic Avatar Projection Overview}. A novel pipeline combining 3D Gaussian Splatting and light-field shading enables real-time, photo-realistic facial avatar hologram animations, streamed to the 3D Looking Glass display. The OpenGL-based light-field shading program uses FLAME expression parameters with its 3D Gaussian primitive correspondence and the display calibration parameters to render a multi-view quilt that adapts to the audience's viewpoint.}
\end{figure*}

Pain assessment is a key skill for healthcare professionals, essential for effective pain management and improving patient outcomes. Traditional training methods using mannequins or classroom materials often fail to reflect real clinical practice, leaving gaps in students' knowledge and skills. Simulated patients provide a more realistic training environment, closely mimicking real-world scenarios. Unlike VR or MR systems, which are limited to individual users, 3D holographic displays allow multiple users to view the same hologram from different angles without headsets. This facilitates group learning and collaborative pain assessment, making it ideal for training. We focus on advanced holographic displays like the Looking Glass, which provide true 3D visualization through volumetric projections. When combined with 3D Gaussian Splatting viewers, these technologies offer flexible, immersive pain recognition training by enabling realistic facial avatars to be observed from various perspectives. Our contributions include:

\begin{enumerate}
    \item We present a proof-of-concept pipeline that connects multimodal input through deep generative modeling to render non-verbal facial expressions via a 3D holographic display. 
    \item We propose solutions to address the challenges inherent in using stereoscopic and holographic displays as simulated patients for pain assessment, focusing on dynamic realism, visual comfort, and interaction fluidity.
\end{enumerate}

\section{Related Work}

\subsection{Simulated Patient}
Unlike classroom materials, which are often theoretical and abstract, the simulated patient approach - having professional certified actors roleplay as a patient in various standardized scenarios, allows trainees to engage with lifelike cases, enhancing their ability to recognize and assess pain in real patients \cite{ma2023standardized}. However, employing human actors requires significant resources to compensate, which leads to the rise of virtual simulated patient development. Mixed reality solutions arise as a cheaper and more sustainable solution in which a virtual avatar is programmed with standardized scenarios for training purposes. Most recent notable mixed reality platforms for nursing training, such as GigXR's Holopatient \cite{kang2023effect} or SimX's VR \cite{shah2023orchestrating}, employ VR headsets as the main medium to project the patient avatar to users. However, due to current technology limitations, VR headset is reported to hinder the training experience: most VR head-mounted display supports only single users, and VR glasses also cause discomfort, isolating feelings with extended use. On the other hand, a multi-view 3D display can naturally support group settings while maintaining realistic interaction between the training and the patient avatar.
\section{System Design}
\subsection{Generative Facial Feedback}
\label{sect:pred}
As the pain-related dataset for simulated patients does not exist, for the first step, we revise the task into a more generalized non-verbal communication behavior endowment for robotic agents. For facial expression, we utilized the FLAME feature \cite{FLAME:SiggraphAsia2017} that represents pose and expression facial features. Pain intensity future may be added into the latent space to adjust facial expression in the future.

Given multi-modal stimulation inputs consist of speaker's audio $A_s(t)$, facial features $V_s(t)$ until time $t$ along with listener's past facial features $V_l(t-k)$ until time $(t-k)$, our model predicts $k$-frames feedback facial features $\hat{V}(t,k)$. We chose 1D convolutional blocks to extract temporal features from modality inputs $F_a, F_v, F_l$.
\begin{align*}
F_a(t) &= F_a(A_s(t)) \\
F_v(t) &= F_v(V_s(t)) \\
F_l(t-k) &= F_l(V_l(t-k))
\end{align*}

The objective is to find the generative model represented by $G$ that predicts $\hat V$ output:
\begin{align*}
    \hat{V}_l(t) &= G\big(F_a(t), F_v(t), F_l(t-k)\big)\\
    H^l_d &= DecoderLayer^l(H^{l-1}_d), H^0_d = X_d(0)
\end{align*}

For this study, we employed the Transformer architecture's Encoder-Decoder modules to decode facial expression feedback from multi-modal encoded embedding. By applying linear transformation $Z$ as the Decoder's last layer, we transformed the decoded embedding into logits for action tokens output:
$$Z(t) = W_{out} H^l_d$$

For every time step in the prediction time window, the highest probability token class was elected as the output: $\hat{y}(t) = \text{argmax}\hspace{0.1cm}{\mathbf{P}(t)}$. The key challenge in facial feedback prediction is fusing multiple modalities (i.e., speaker's audio/visual and listener's visual) into one context embedding. Given two input modalities $X$ and $Y \in \mathbb{R}^{T\times D}$, by concatenation along feature dimension $D$ we have:
$$
F_{concat}(t) = \text{concat}(X(t),Y(t))
$$

\begin{figure*}[t!]
\centering
\includegraphics[width=1\textwidth, trim={1cm 0.5cm 1cm 0.5cm}, clip]{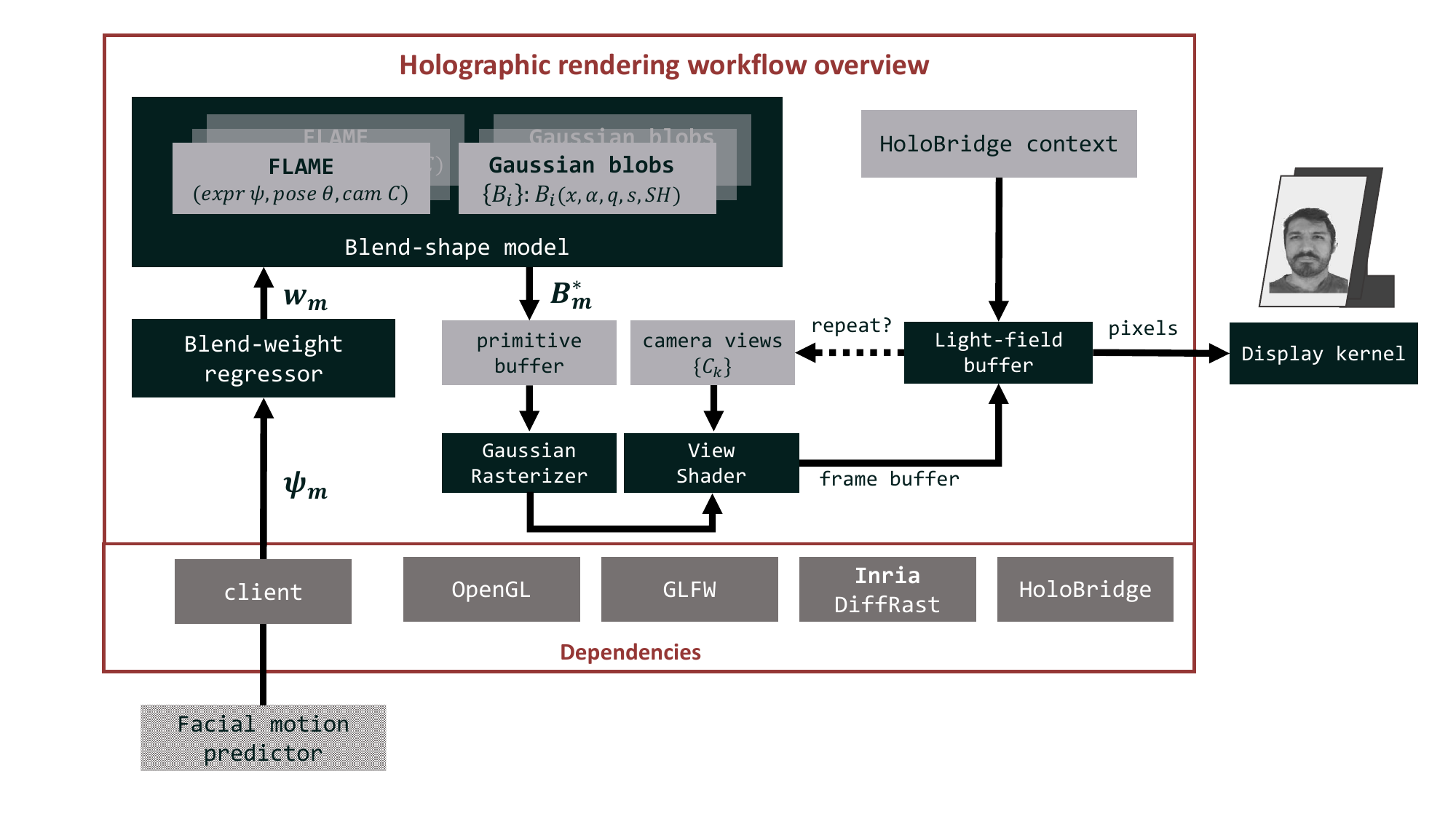}
\caption{\label{fig:offset_scheme}Holographic rendering pipeline overview. Our system consists of a blend-shape model for the 3D Gaussian Splatting Avatar with a FLAME prior. Given an expression parameter $\psi_m$ at frame $\#m$, the pipeline linearly blends pre-trained blend shapes ${B_K}$ into an accumulated $B^*_m$ primitive parameters. An iterative procedure rasterizes this 3D facial avatar from various viewpoints to reproduce a light field given an input viewpoint. The final output is a shaded quilt comprised by multiple discrete views that visualizes the expected 3D hologram on the target 3D display.}
\end{figure*}

By transforming the concatenated features into learned query, key, and value matrix $W_q$, $W_k$, and $W_v$, attention weight can be computed to add create cross-relationship to each modality's time step, creating fused embedding $C(t)$:
$$C(t) = \text{softmax}\big(\frac{QK^T}{\sqrt{d_k}}\big) V$$

\begin{algorithm}
\caption{Calibrate Multi-View 3D Avatar}
\label{alg:calib}
\begin{algorithmic}[1]
\REQUIRE Virtual camera parameters $camSize$, aspect ratio $ar$, field-of-view $fov(.)$, viewer's distance $d$, base view matrix $V$
\ENSURE Multi-view quilt $views$
\STATE Initialize empty array $views$[ ]
\STATE $total \gets$ 48
\STATE $fov_d \gets $fov$(d)$
\STATE $d_{cam} \gets \frac{-camSize}{\tan(fov_d / 2)}$
\FOR{$i \gets 0$ to $(total - 1)$}
\STATE $\alpha_{off} \gets alpha_{max} * (\frac{i}{total - 1} - 0.5)$
\STATE $t_{off} \gets d_{cam} *\tan(alpha)$
\STATE
\STATE \# Compute view matrix
\STATE $V'\gets$Translate($V,t_{off}, camDist$)
\STATE
\STATE \# Projection matrix for parallel projection.
\STATE $P \gets$ Perspective($fov, ar, z_{near}, z_{far}$)
\STATE $P[2][0]\gets P[2][0]+t_{off}/(camSize * ar)$
\STATE
\STATE \# Render 2D portrait
\STATE $v \gets$ render($V'$, $P$)    
\STATE Append $v$ to $views$[ ]
\ENDFOR
\STATE \RETURN $views$[ ]
\end{algorithmic}
\end{algorithm}

Finally, by splitting feature dimension now $(D_{X} + D_{Y})$ by $D_X$ we had a new attention-aware embedding of $X$ fused by $Y$.

\begin{figure}[t]
\centering
\includegraphics[width=0.5\textwidth, trim={7cm 8cm 7cm 0cm}, clip]{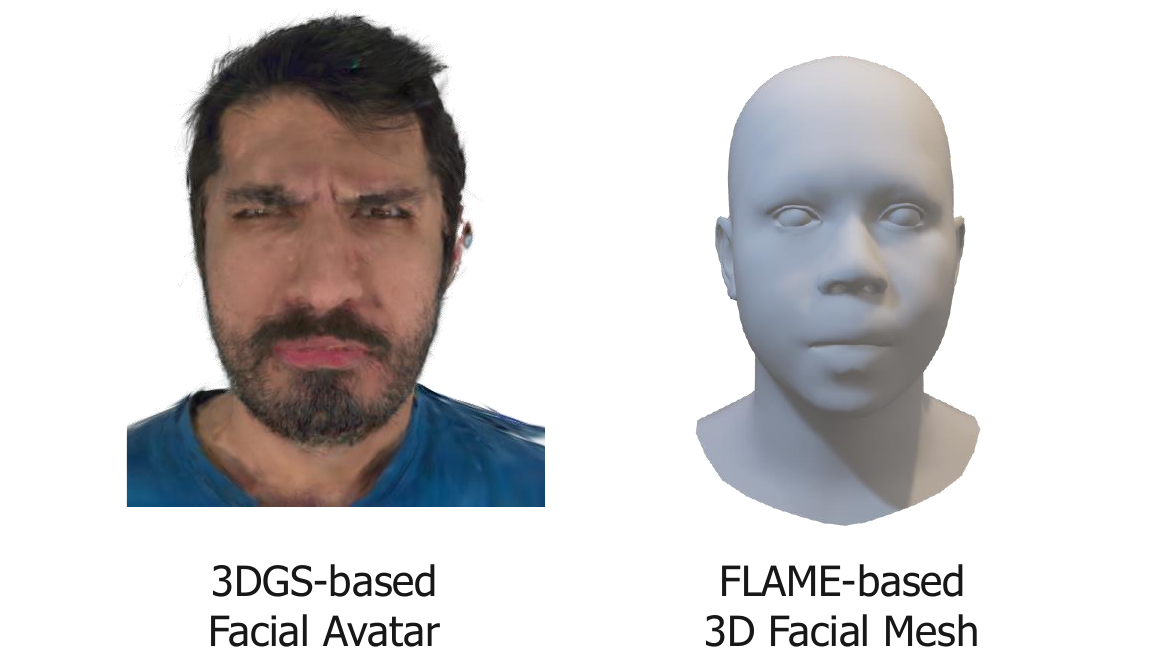}
\caption{\label{fig:FLAME}3D FLAME output features could be used to reconstruct 3D facial mesh model that works as prior for photorealistic 3D Gaussian Splatting facial avatar.}
\end{figure}

\begin{table*}[!t]
\centering
\caption{\textbf{Baselines.} Comparison of our approach with baseline models \cite{song2024react,damfinite} on the test set.}
\begin{tabular}{lllllll} 
\toprule
                                               & \multicolumn{2}{l}{\textbf{Appropriateness}}                  & \multicolumn{2}{l}{\textbf{Diversity}}                    & \textbf{Realism}              & \textbf{Synchrony}        \\ 
\cmidrule(lr){2-7}
                                               & \textbf{FRCorr} ($\uparrow$) & \textbf{FRDist} ($\downarrow$) & \textbf{FRDiv} ($\uparrow$) & \textbf{FRVar} ($\uparrow$) & \textbf{FRRea} ($\downarrow$) & \textbf{FRSyn} ($\cdot$)  \\ 
\midrule
Ground truth                                   & 8.73                         & 0.00                           & 0.0000                      & 0.0724                      & -                             & 47.69                     \\ 
\midrule
Random                                         & 0.05                         & 237.23                         & 0.1667                      & 0.0833                      & -                             & 44.10                     \\
Mime                                           & 0.38                         & 92.94                          & 0.0000                      & 0.0724                      & -                             & 38.54                     \\
MeanFr                                         & 0.00                         & 97.86                          & 0.0000                      & 0.0000                      & -                             & 49.00                     \\ 
\midrule
Trans-VAE\cite{song2024react}                                      & 0.07                         & 90.31                          & 0.0064                      & 0.0012                      & 69.19                         & 44.65                     \\
BeLFusion\cite{song2024react}                                      & 0.12                         & 94.09                          & 0.0379                      & 0.0248                      & 94.09                         & 49.00                     \\
FSQ-Tformer\cite{damfinite}                                    & 0.31                         & \textbf{84.93}                 & 0.1164                      & 0.0348                      & 34.66                         & \textbf{47.42}                     \\
\rowcolor[rgb]{0.855,0.91,0.988} \textbf{Ours} & \textbf{0.52}                & 86.23                          & \textbf{0.1212}             & \textbf{0.0442}             & \textbf{25.78}                & 48.57            \\ 
\bottomrule
                                               &                              &                                &                             &                             &                               &                           \\
\multicolumn{7}{l}{($\cdot$) means the closer to the ground truth, the better.}                                                                                                                                                        \\
\multicolumn{7}{l}{indicates the best average performance among the heuristic baselines for the groups of metrics.}                                                                                                                   
\end{tabular}
\label{tab:resulttable}
\end{table*}

\subsection{3D Facial Avatar Rendering}

In current facial avatar work, 2D rendering from generative model outputs is common. However, our research explores converting these outputs into 3D facial avatars compatible with multi-view stereoscopic displays like the Looking Glass Portrait (LGP).

\subsubsection{3D Gaussian Splatting Avatar}

Inspired by recent work on 3D Gaussian Splatting Avatar \cite{qian2023gaussianavatars}, we adapt this implementation into our current task to render expression transition from our predictor in Section \ref{sect:pred}. The main advantages of 3D GaussianAvatar include its extremely fast rendering speed ($\ge 60$fps, tested on Nvidia RTX 4060 GPU). Given a FLAME expression/pose facial codes feature along with camera perspective, and a pre-trained avatar model, GaussianAvatar can render a realistic 2D portrait. After training, GaussianAvatar reconstructs a rigged FLAME-based 3D Gaussian Splatting avatar \cite{qian2023gaussianavatars} defined as:
\begin{align}
\mathcal{F}_{\text{3DGS}}(\mathbf{p}, \mathbf{M})\quad\textrm{where } &&\mathbf{p}&: \textrm{3DGS primitives}\\
&&\mathbf{M}&: \textrm{FLAME mesh}
\end{align}

Driven by the expression code output at Sect. \ref{sect:pred}, the 3D facial FLAME mesh (right Fig.\ref{fig:FLAME}) morphed from one 3D face shape to another along with the Gaussian incorporated with the original FLAME mesh's face elements.

\subsubsection{Multi-view Display Calibration}

LGP collects 48 discrete views of a given 3D scene to project them across a 40-degree-wide view cone. The 48-view input requires a calibration function $\Phi_C$ based on the display's optical properties:
$$
\Phi_C(\mathbf{fov}, \mathbf d, \mathbf \alpha_{off}, \mathbf{t}_{off}, \mathbf{ar})
$$
The $\Phi_C$'s main objective is to extract the perspective view and projection matrix from situational parameters such as display device properties (aspect ration $\mathbf{ar}$, field-of-view $\mathbf{fov}$) and user's viewing position parameters (angle $\mathbf\alpha$, distance $\mathbf D$) by calculating offset parameters (translation $\mathbf{t}_{offset}$ and rotation $\mathbf \alpha_{offset}$, see Fig.\ref{fig:offset_scheme}) to correct the projection of neighbor views merged into each other at audience's perspective. For simplification, we assume the discrete view is distributed horizontally on which the audience will observe the stereoscopic display. The detailed process was described in Alg.\ref{alg:calib} where offset translation was calculated before being applied to the view matrix and projection matrix for each discrete view. The process is repeated with 48 views before sending to the LGD device via USB interface. 

\section{Evaluation}

We evaluate the current generative facial expression according to existing quantitative metrics regarding appropriateness, diversity, realism, and synchrony \cite{song2024react}, see Table \ref{tab:resulttable}. The adopted method showed improvement across various metrics that could verify the quality of generated non-verbal reactions given multi-modal stimulation from a conversation. In the future, we plan to translate the current non-verbal facial reaction taking into account pain intensity for the pain assessment training as mentioned at the beginning.

\section{Conclusion}

Integrating photo-realistic interactive avatars into medical training for pain assessment holds great promise for enhancing the skills of healthcare professionals. By focusing on technological enhancements, improving interactivity, conducting rigorous validation, and ensuring scalability, this research can significantly contribute to more effective and accessible medical education.

% \vfill\eject
% \small
\bibliographystyle{plain}
\bibliography{refs}

% \printbibliography

% \begin{thebibliography}{9}
% %%%%%%%%%%%%%%%%%%%%%%%%%%%%%%%%%%%%%%%%%%%%%%%%%%%%%%%%%%%%%%%%%%%%%%%%%%%%%%%

% \bibitem{website}
% ``International Session web site for the 42nd Annual Conference of the Robotics Society of Japan,''
% \url{ https://ac.rsj-web.org/2024/is.html}
% \bibitem{yamada2000}
% Taro Yamada and Ichiro Suzuki:
% ``A New Style of Manuscript for 100th Annual Conference of the Robotics Society of Japan,''
% Journal of the Robotics Society of Japan, Vol. 100, No. 4, pp. 8--12, 2082.

% %%%%%%%%%%%%%%%%%%%%%%%%%%%%%%%%%%%%%%%%%%%%%%%%%%%%%%%%%%%%%%%%%%%%%%%%%%%%%%%

\normalsize
\end{document}